\documentstyle[aps,prb,twocolumn,epsfig,floats,amssymb,amsfonts]{revtex} 
 
\def\s{\sum\limits}  
\def\p{\prod\limits}  
\def\pa{\partial}  
\def\i{\int\limits}  
\def\be{\begin{equation}}  
\def\e{\end{equation}}  
\def\beml{\begin{mathletters}}  
\def\eml{\end{mathletters}}  
\def\beq{\begin{eqnarray}}  
\def\eq{\end{eqnarray}}  
\def\ba{\begin{array}}  
\def\a{\end{array}}  
\def\d{\dagger}  
\def\l{\left}  
\def\r{\right}  
\def\n{\nonumber}  
\def\la{\langle}  
\def\ra{\rangle}  
\def\det{\,{\rm Det}\,}

\def\im{\,{\rm Im}\,}  
\def\re{\,{\rm Re}\,}  
  
\begin{document}  
   
\draft  
  
\date{January 2001}  
  
\title{Fokker-Planck equations and density of states in disordered   
quantum wires}  
  
\author{M.\ Titov$^{a}$, P. W. Brouwer$^{b}$, A. Furusaki$^{c}$,  
C. Mudry$^{d}$}   
\address{  
{}$^a$Instituut-Lorentz, Universiteit Leiden,  
P.\,O.~Box 9506, 2300 RA Leiden,  
The Netherlands\\  
{}$^b$Laboratory of Atomic and Solid State Physics,  
Cornell University, Ithaca, NY 14853-2501\\  
{}$^c$Yukawa Institute for Theoretical Physics,   
Kyoto University,  
Kyoto 606-8502, Japan\\  
{}$^d$Paul Scherrer Institut, CH-5232 Villigen PSI, Switzerland  
}  
\twocolumn[  
\widetext  
\begin{@twocolumnfalse}  
  
\maketitle  
  
\begin{abstract}  
We propose a general scheme to construct scaling equations for  
the density of states in disordered quantum wires for all ten  
pure Cartan symmetry classes.  
The anomalous behavior of the density of states near the Fermi level  
$\varepsilon=0$ for the three chiral and four Bogoliubov-de Gennes   
universality classes is analysed in detail by means of a  
mapping to a scaling equation for the reflection from a quantum  
wire in the presence of an imaginary potential.    
\end{abstract}  
  
  
\end{@twocolumnfalse}  
]  
  
\narrowtext  
  
\section{Introduction}  
  
Statistical properties of energy levels and wavefunctions in  
disordered electron systems are believed to be determined by,  
first of all, the fundamental symmetries of the Hamiltonian. Such  
a connection is best established in the case of ``zero-dimension'',  
i.e., for finite size systems with a large dimensionless   
conductance $g$,  
where a description in terms of random matrix theory is valid.  
Using the link to Cartan's classification of symmetric   
spaces,\cite{Helgason} Zirnbauer\cite{Zirnbauer} and   
Caselle\cite{Caselle}   
have pointed out that there exist only ten possible random matrix   
theories, whose form follows   
directly from the geometrical characteristics (``roots'') of   
the corresponding symmetric space in Cartan's table.   
These ten random matrix theories are divided into three  
standard classes,\cite{Mehta}  
three chiral classes,\cite{Chiralclasses}   
and four Bogoliubov-de Gennes (BdG) classes,\cite{AZ}   
the subdivision in each class depending on   
the presence or absence of time-reversal symmetry (TR) and   
spin-rotation invariance (SR),\cite{foot} see Table \ref{tab:1}.   
  
The rational for the Cartan classification is believed to  
transcend ``zero-dimension''  
and has been applied to the construction of effective theories   
(e.g., non-linear-sigma-models) for higher dimensional disordered systems.  
The standard classes are thus believed to be appropriate to the problem of an  
electron moving in a random potential, without further  
symmetries,\cite{Anderson1958} irrespective of dimensionality.  
The chiral classes  
are appropriate to the case when the disorder is purely   
``off-diagonal'',\cite{Gade}   
as is the case, e.g., for the lattice random flux model  
in quasi-one- and two-dimensions,\cite{RF,Mudry1999}   
the random hopping model,\cite{real-pi-flux,Fabrizio2000},  
and for random XY spin chains.\cite{randomXY}   
The BdG classes refer to systems with superconducting correlations,\cite{ASZ}  
and were argued to be valid for   
vortices in superconductors,\cite{Skvortsov1997,Bundschuh1998}  
dirty unconventional superconductors,\cite{Senthil98,unitary}   
and (in the case of broken time-reversal symmetry)   
normal metals in proximity to a superconductor.  
The BdG classes have also been argued to be of some  
relevance to problems in statistical mechanics such as  
random bond Ising and network models.\cite{rbim,network}  
  
\begin{figure}  
\begin{center}  
\begin{picture}(200, 30)(0,10)  
\put(  0,22){({\rm a})}            \put(110,22){({\rm b})}  
\put(  0, 7){$r(\varepsilon)$}            \put(110,7){$r(\varepsilon)$}  
\thicklines  
\put( 20, 0){\line(1,0){80}} \put(130, 0){\line(1,0){80}}  
\put( 20,20){\line(1,0){80}} \put(130,20){\line(1,0){80}}  
\put( 20, 5){$ \leftarrow$}  \put(130, 5){$ \leftarrow$}  
\put( 20,10){$\rightarrow$}  \put(130,10){$\rightarrow$}  
\thicklines  
\put( 30, 0){\line(0,1){20}} \put(140, 0){\line(0,1){20}}  
\put( 90, 0){\line(0,1){20}} \put(210, 0){\line(0,1){20}}  
                             \put(209.5,0){\line(0,1){20}}  
                             \put(209  ,0){\line(0,1){20}}  
\thinlines  
\multiput( 30, 20)(5,0){9}{\line(1,-1){20}}  
                             \multiput(140,  
20)(5,0){11}{\line(1,-1){20}}  
\put( 30, 15){\line(1,-1){15}}  
\put( 30, 10){\line(1,-1){10}}  
                             \put(140, 15){\line(1,-1){15}}  
                             \put(140, 10){\line(1,-1){10}}  
\put( 90,  5){\line(-1,1){15}}  
\put( 90, 10){\line(-1,1){10}}  
                             \put(210,  5){\line(-1,1){15}}  
                             \put(210, 10){\line(-1,1){10}}  
  
\end{picture}  
\hfill\break  
\end{center}  
\caption{\label{fig:wire}  
A quantum wire (hashed marked region) of length $L$ connected to ideal leads.  
The reflection matrix at energy $\varepsilon$, $r(\varepsilon)$,   
is the matrix of reflection amplitudes  
for reflection into the same lead.  
(a) A two lead geometry is used to compute the conductance.  
(b) A one lead geometry is used to compute the DOS whereby the wire  
is closed by a perfectly reflecting wall on the right.  
}  
\end{figure}
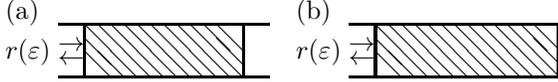

The geometric structure of the symmetric spaces not only determines  
the level statistics in zero-dimension, it also determines the  
form of scaling equations for the transmission eigenvalues $T_j$  
and reflection eigenvalues $R_j = 1-T_j$   
($j=1,\ldots,N$) in a  wire geometry,   
where a disordered sample of length $L$ is connected to two ideal leads   
(see Fig.~\ref{fig:wire}a).\cite{Caselle,Hueffmann,Beenakker}   
Here the reflection eigenvalues $R_1,\dots, R_N$  
are the $d$-fold degenerate eigenvalues of the matrix  
product $r^\d r$, where $r$ is the $Nd$-dimensional matrix  
of reflection amplitudes for back-reflection into the same lead  
(see Fig.\ \ref{fig:wire}a)  
and $Nd$ is the number of propagating channels at the Fermi level.  
The degeneracies $d$ are listed in the Table\ \ref{tab:1}  
for the 10 symmetry classes in Cartan's classification.  
Parameterizing  
\begin{equation}  
  R_j = \tanh^2 x_j,   
\label{eq:R}  
\end{equation}  
the scaling equation for the distribution $P(x_1,\ldots,x_N;L)$  
takes the form \cite{Caselle,exception}  
\begin{eqnarray}  
  {\partial P \over \partial L} &=&  
  {1 \over 2 \gamma \ell} \sum_{j=1}^{N} {\partial \over \partial x_j} J   
  {\partial \over \partial x_j} J^{-1} P,  
  \label{eq:PL}  
\end{eqnarray}  
where $\ell$ is the mean free path and $L$ is the length of the quantum wire.  
For the standard and BdG symmetry classes, the $x_j$ are positive  
random variables, $0 < x_j < \infty$, and the Jacobian $J$   
and the numerical constant $\gamma$ read  
\beml  
\label{J1}  
\begin{eqnarray}  
  J &=& \prod_{j=1}^{N} \sinh^{m_l}(2 x_j)   
      \prod_{k < j} \prod_{\pm}  
   \sinh^{m_o}(x_j\pm x_k),\\  
   \gamma&=& m_o(N-1)+m_l+1,  \label{eq:gammaBdG}  
\end{eqnarray}  
\eml  
while for the chiral classes the $x_j$ can take values on   
the entire real axis, $-\infty < x_j < \infty$, and one has  
\beml  
\label{J2}  
\beq  
  J &=& \prod_{j=1}^{N}   
      \prod_{k < j} \sinh^{m_o}(x_j - x_k),\\  
  \gamma&=&\case{1}{2}\l[m_o(N-1)+2\r]. \label{eq:gammachiral}  
\eq  
\eml  
Here $m_l$ and $m_{o}$ are the multiplicities of the   
so-called long and ordinary roots for the corresponding symmetric  
space,\cite{footmo} respectively, see Table \ref{tab:1}. (For  
the chiral classes we only consider the case when the  
nonuniversality parameter $\eta$ equals unity, see Ref.\  
\onlinecite{BMFNucPhB}.)  
For the standard classes, Eq.~(\ref{eq:PL}) was derived by Dorokhov,  
\cite{Dorokhov} and Mello, Pereyra, and Kumar;\cite{MPK}  
derivations for the chiral and BdG symmetry classes can be found  
in Refs.~\onlinecite{BMSA} and \onlinecite{BFGM}.  
  
In this note, we show how the same structure also determines   
scaling equations for the density of states (DOS)  
$\nu(\varepsilon)$ in a quantum wire of infinite length   
($L\to\infty$) at energy $\varepsilon$.   
Our work builds on previous work for the chiral classes, where such a  
scaling equation was derived using a different method.\cite{BMF}  
In addition, we present an exact solution for the DOS,  
something that could not be done in Ref.~\onlinecite{BMF}.   
  
While $\nu(\varepsilon)$ is a non-singular function of energy  
for the three standard classes, singular behavior is expected  
near $\varepsilon=0$ for the remaining seven symmetry classes,  
as $\varepsilon=0$ is a special point there. (The energy  
$\varepsilon=0$ corresponds to a point of particle-hole symmetry  
in the chiral and BdG symmetry classes; it is the band center for  
lattice models with random hopping, or the Fermi energy in the  
case of the BdG classes.) Indeed, for  
the chiral classes, $\nu(\varepsilon)$ was found to depend   
sensitively on the parity of the channel number  
$N$ for $\varepsilon$ close to zero, showing the   
$\nu(\varepsilon) \propto 1/|\varepsilon\ln^3 \varepsilon|$   
divergence characteristic of pure one-dimensional  
systems with chiral symmetry for odd $N$,\cite{Dyson53} while  
$\nu(\varepsilon) \propto |\varepsilon^{m_l-1} \ln \varepsilon|$  
for even $N$.\cite{BMF}   
For the BdG classes in the presence of spin-rotation  
invariance (classes C and CI; we refer to  
the symmetry classes by Cartan's symbol for   
the symmetric space corresponding to their Hamiltonians, see  
Table \ref{tab:1}),   
a suppression $\nu(\varepsilon) \propto \varepsilon^{m_l-1}$  
is expected.\cite{Senthil98,Senthil99}   
In both cases, the characteristic energy  
scale for the DOS singularity is $\sim \hbar v_F/N^2 \ell$, $v_F$  
being the Fermi velocity. This distinguishes the singularity in   
$\nu(\varepsilon)$ for quantum wires from its counterpart in   
zero-dimension,   
where the characteristic energy scale is the mean  
level spacing, which goes to zero as the system size is increased.  
No exact results are known for the multichannel quantum wires  
from BdG symmetry classes in the absence of spin-rotation invariance  
(classes D and DIII;  
see however Refs.~\onlinecite{Motrunich00,GruzbergPriv,Senthil00}   
 for asymptotic results in one- and two-dimensions).   
The general scheme that we present here fills this gap, and provides a   
unified framework for all ten symmetry classes.\cite{foot2}  
  
\begin{table}  
\begin{tabular}{l|l|l|l|l|l|l|l}  
class & TR & SR & $m_o$ & $m_l$ & $d$ & ${\cal M}$ & ${\cal H}$ \\ \hline  
& Yes & Yes & 1 & 1 & 2 & CI& AI \\  
standard & No & Yes(No)& 2 & 1 & 2(1)& AIII& A \\  
& Yes & No & 4 & 1 & 2 & DIII& AII \\ \hline   
& Yes & Yes & 1 & 0 & 2 & AI& BDI \\  
chiral & No & Yes(No)& 2 & 0 & 2(1) & A& AIII \\  
& Yes & No & 4 & 0 & 2 & AII& CII \\ \hline   
& Yes & Yes & 2 & 2 & 4 & C& CI \\  
BdG & No & Yes & 4 & 3 & 4 & CII& C  \\  
& Yes & No & 2 & 0 & 2 & D& DIII \\  
& No & No & 1 & 0 & 1 & BDI& D   
\end{tabular}\medskip  
  
\caption{\label{tab:1} Classification of symmetry classes according  
to Cartan's table. The symmetry classes are defined in terms of the   
presence or absence of  
time-reversal symmetry (TR) and spin-rotation invariance (SR), and  
in terms of the other fundamental symmetries of the system  
(standard, chiral, or BdG). The table further contains the  
multiplicities of the   
ordinary and long roots $m_o$ and $m_l$ and the degeneracy  
$d$ of the reflection/transmission eigenvalues.\protect\cite{footmo}   
In the seventh column we denote, using Cartan's notation  
(chapter X of Ref.~\protect\onlinecite{Helgason}),  
the symmetric spaces associated to the transfer matrix group   
${\cal M}$ of the quantum wire; these are different from the symmetric spaces   
associated to the microscopic Hamiltonian ${\cal H}$ of the quantum wire,  
which are listed in the last column. Following the convention   
of Ref.~\protect\onlinecite{AZ}, we refer to the 10 symmetry classes   
by Cartan's symbol for the symmetric space of their Hamiltonians.  
For standard and chiral quantum wires without TR  
the degeneracy $d$ of the reflection eigenvalues is 2 (1) with   
(without) SR.}  
\end{table}  
  
\section{Mapping to quantum wires with absorption}  
  
Our construction of a scaling equation describing the DOS  
makes use of a mapping to a quantum wire with absorption.  
Absorption is described by the addition of a spatially uniform   
imaginary potential $i \omega$, $\omega>0$,  
to the Hamiltonian.   
In the presence of absorption, a scaling equation for the   
reflection eigenvalues  
$R_j$, $j=1,\ldots,N$,  
 of $r(i\omega)^{\dagger} r(i \omega)$ can be derived in the  
same way as the scaling equation (\ref{eq:PL}) in the absence of  
absorption,\cite{Bee96,Chalker} noting  
that upon addition of a thin slice of length $\delta L$ to a  
disordered wire of length $L$, the reflection eigenvalues  
$R_j$ change by a small   
amount $\delta R_j$. The change $\delta R_j$ comes from   
two {\em statistically independent} contributions:  
(1) disorder scattering in the added slice of width $\delta L$,  
and (2) absorption in the same slice. These contributions are   
statistically independent, since to linear order in $\delta L$ one can   
neglect processes that   
involve both disorder scattering and absorption. The second   
process gives rise to a change  
\be  
  R_j \to e^{-4\omega \delta L/v_F} R_j  
\qquad j=1,\ldots,N,  
\e  
independent of the disorder. The first process depends on the   
disorder type and symmetries, and is already present in the  
scaling equation (\ref{eq:PL}). Combination of the two processes  
then results in the replacement of Eq.~(\ref{eq:PL}) by  
\be  
\label{FP}  
\frac{\pa P_{i\omega}}{\pa L}=  
  \s_{j=1}^{N} {\pa \over \pa x_j}  
  \left( {\omega \over v_F} \sinh 2 x_j +  
{1 \over 2 \gamma \ell}  J   
  \frac{\pa}{\pa x_j} J^{-1} \right) P_{i\omega},  
\e  
where we used the parametrisation (\ref{eq:R}) for the $R_j$,  
and $P_{i\omega}(x_1,\ldots,x_N;L)$ represents the joint distribution of the  
reflection eigenvalue in the presence of the imaginary potential  
$i \omega$.   
Equation (\ref{FP}) was derived in Refs.~\onlinecite{Bee96,Chalker}   
for  
the standard symmetry classes. The initial condition at $L=0$  
for $P_{i\omega}(x_1,\dots, x_N;L)$ still needs to be specified.  
For our purposes, it is advantageous to set    
$R_j = 1$, $j=1,\ldots,N$, at $L=0$,  
corresponding to a wire that is closed on one  
end, see Fig.~\ref{fig:wire}b.  
  
The scaling equation (\ref{FP}) has the stationary   
(independent of $L$) solution,  
\be  
\label{solution}  
P_{i\omega;{\rm st}}(x_1,\ldots,x_N)  
=\frac{1}{Z(a)} |J| \p_{j=1}^{N}   
  e^{-a \cosh{2 x_j}},  
\e  
where $a = \gamma \ell \omega/v_F$ and the normalization constant  
$Z(a)$ ensures that  
the stationary solution $P_{i\omega;{\rm st}}$ is normalized to one.   
This solution represents the joint probability density of the   
reflection coefficients for the absorbing medium. We will demonstrate below  
that the DOS in an infinitely long disordered wire without  
absorption can be computed as a logarithmic derivative  
of the normalization constant or ``partition function''  
 $Z(a)$ with respect to the   
dimensionless imaginary energy $a$.  
  
The main idea of our method is to make the analytical continuation   
$i\omega \to \varepsilon$. With this analytical continuation, the  
single-energy product $r(i\omega)^{\dagger} r(i \omega)$ of reflection  
matrices in the  
presence of absorption maps to  
the two-energy product $r(-\varepsilon)^{\dagger} r(\varepsilon)$ without  
absorption.\cite{BB} The latter product is a unitary matrix  
since the reflection matrix in the geometry of Fig.~\ref{fig:wire}b  
is unitary by construction.  
The $N$ independent eigenvalues of $r(-\varepsilon)^{\dagger} r(\varepsilon)$  
can thus be written as   
$\exp(2 i \phi_j)$, $j=1,\ldots,N$. When the length $L$ of the disordered  
wire is increased, the $\phi_j$'s perform a Brownian motion, governed  
by a Fokker-Planck equation that is the analytical continuation of   
Eq.~(\ref{FP}),  
\begin{mathletters} \label{eq:FPphiall}  
\be  
\label{eq:FPphi}  
  {\partial P_{\varepsilon} \over \partial L} =  
  \sum_{j=1}^{N}  
  {\partial \over \partial \phi_j}  
  \left( -{2\varepsilon \over v_F} + {2 \over \gamma \ell}  
  \sin^2 \phi_j J {\partial \over \partial \phi_j}  
  J^{-1} \right) P_{\varepsilon},  
\e   
where   
\begin{equation}  
  J=\p_{j=1}^{N} {1 \over \sin^{\gamma} \phi_j}  
    \p_{k<j} \sin^{m_o}(\phi_{j} - \phi_{k})  
\end{equation}  
for the standard and BdG classes, and  
\begin{equation}  
  J=\p_{j=1}^{N} {1 \over \sin^{\gamma} \phi_j}  
    \p_{k<j} \sin^{m_o} [(\phi_{j} - \phi_{k})/2]  
\end{equation}  
\end{mathletters}  
for the chiral classes, where $\gamma$ is defined in Eqs.\  
(\ref{J1}) and (\ref{J2}), respectively.  
This equation was derived by different methods in  
Ref.~\onlinecite{BMF} for the chiral classes and in  
Ref.~\onlinecite{TB} for the standard classes.   
The scaling equation (\ref{eq:FPphiall}) can be considered either 
on the full real axis $-\infty < \phi_j < \infty$, or 
on the interval $-\pi \le \phi_j < \pi$. In the latter case   
Eq.~(\ref{eq:FPphiall}) has a well-defined stationary solution   
$P_{\varepsilon;{\rm st}}(\{\phi_{j}\})$ describing the distribution   
$P_{\varepsilon}(\{\phi_{j}\};L)$ in the  
limit $L \to \infty$. However, this solution is not of the simple form  
(\ref{solution}).\cite{footst} In fact, no representation of 
$P_{\varepsilon;{\rm st}}$ in terms of elementary functions 
could be found other than for the case $N=1$, where  
explicit integration of Eq.~(\ref{eq:FPphiall}) is possible.\cite{OE}   
 
The DOS is calculated via the ``node-counting theorem'', which states 
that, for the half-open quantum wire of Fig.\ \ref{fig:wire}b, the 
mean number of states ${\cal N}(\varepsilon)$ 
in the energy interval $(-\varepsilon,\varepsilon)$ and per unit 
length is proportional to  
the sum of the phase derivatives  
$\partial \phi_j/\partial L$.\cite{Schmidt,But93} For 
technical reasons, we replace $\partial \phi_j/\partial L$  
by the derivative of the counting function  
$-{\rm Im}\,\ln\sin(\phi_j+i0)$, which has the same average 
slope as $\phi_j$. We thus find 
\begin{eqnarray}  
  {\cal N}(\varepsilon) &\equiv& 
  \int_{-\varepsilon}^{\varepsilon} d\varepsilon' \nu(\varepsilon') 
\nonumber\\ 
&=& 
 - {d \over \pi} \s_{j=1}^N  {\partial\over \partial L}  
  \left\langle \im \ln \sin (\phi_j + i0) \right \rangle_{\varepsilon,L}. 
\label{eq:cal N}  
\end{eqnarray}  
Note that the positive infinitesimal $i0$ in the argument  
of $\sin (\phi_j + i0)$, which is needed to select the correct branch of  
the logarithm, 
is compatible with the analytical continuation $i \omega \to \varepsilon$  
used to obtain the scaling equation (\ref{eq:FPphiall}). 
In Eq.\ (\ref{eq:cal N}), the brackets 
$\langle\ldots\rangle_{\varepsilon,L}$ denote  
an average with the probability density 
$P_{\varepsilon}$ that solves Eq.~(\ref{eq:FPphiall}) on the full real 
axis $-\infty < \phi_j < \infty$ at length $L$ of the quantum wire. 
Integrating by parts, with the help of Eq.\ (\ref{eq:FPphiall}), we find  
\begin{eqnarray}  
  {\cal N}(\varepsilon) &=& -{d \over \pi}  
  \sum_{j=1}^{N} \int d\phi_1 \ldots d\phi_N  
  \im \left[\cot (\phi_j + i0)\right]  
  \nonumber \\ && \mbox{} \times  
  \left(  
  {2\varepsilon \over v_F}  
  - {2\over\gamma\ell}\sin^2\phi_jJ{\partial \over \partial \phi_j}J^{-1}  
  \right) P_{\varepsilon} 
  \nonumber \\  
  &=&  
  - {2d\varepsilon\over \pi v_F} \sum_{j=1}^{N}  
  \langle \im \cot(\phi_j + i 0)\rangle_{\varepsilon}.  
\end{eqnarray}  
Now, in the thermodynamic limit $L \to \infty$, the 
disorder average $\langle \ldots 
\rangle_{\varepsilon}$ 
can be performed with   
the stationary distribution $P_{\varepsilon;{\rm st}}$  
for the interval $-\pi \le \phi_j < \pi$.  
Hence the DOS per unit length is given by  
\begin{eqnarray}  
\nu (\varepsilon) =  
{1 \over 2} \frac{\pa {\cal N}}{\pa \varepsilon} 
= 
-\frac{d}{\pi v_F} 
  \sum_{j=1}^{N}  
\frac{\pa}{\pa \varepsilon}  
\varepsilon  
\left\langle \im \cot (\phi_j + i0) \right\rangle_{\varepsilon} 
,  
\label{DOS}  
\end{eqnarray}  
where we have used the relation   
\begin{equation}  
\nu(\varepsilon)= \nu(-\varepsilon)  
\label{eq: chirality on spec}  
\end{equation}  
that holds for each disorder configuration in the chiral and BdG  
classes. In the standard classes, Eq.~(\ref{eq: chirality on spec})  
only holds in the thermodynamic limit and for sufficiently small   
$\varepsilon$ (smaller than the energy scale $v_F/\ell$, where  
nonuniversal effects appear and the validity of the scaling   
equations breaks down).  
  
Instead of performing the disorder average with the   
stationary solution $P_{\varepsilon;{\rm st}}$  
of Eq.~(\ref{eq:FPphiall}), we rely on the solution  
$P_{i\omega;{\rm st}}$ for the stationary distribution of the  
reflection eigenvalues in the presence of the imaginary   
potential $i \omega$, see Eq.\ (\ref{solution}).   
From the relation $\tanh^2 x_j =e^{2 i \phi_j}$ 
we first observe that  
\be  
  \im \langle \cot (\phi_j + i0) \rangle_{\varepsilon} = - 
  \re \left. \langle \cosh 2x_j \rangle_{i\omega} 
  \right|_{\omega \to -i \varepsilon},  
\e  
where   
$\l\la \ldots\r\ra_{i\omega}$  
stands for averaging with the stationary distribution  
$P_{i\omega;{\rm st}}$.  
Second, we note that the average   
$-\sum_m\la\cosh 2x_m\ra_{i\omega}$ is given by   
the logarithmic derivative of the partition function   
$Z(a)$, thus obtaining the DOS  
\be  
\label{main}  
\nu(\varepsilon)=  
-\frac{d}{\pi v_F}  
\re\left. \frac{\pa}{\pa a}  
\left[a\frac{\pa}{\pa a}  
\ln{Z(a)}  
\right] \right|_{a\to -i \gamma \ell\varepsilon /v_F}.  
\label{eq: master eq}  
\e  
This relation between the DOS   
and the partition function $Z(a)$  
for each of the Cartan symmetry classes  
is the key result of this paper.  
  
\section{Calculation of the density of states}  
  
As we have seen in the previous section, the computation of the DOS  
in the thermodynamic limit reduces to the computation of  
the ``partition function''  
\begin{equation}  
Z(a)=  
\int dx_1\ldots dx_N\,  
|J(\{x_j\})|\, e^{-a\sum_{j=1}^N\cosh 2x_j}.  
\label{eq:Zx}  
\end{equation}  
For the standard and BdG classes, the integration over the  
coordinates $x_j$ is restricted to the positive real axis,  
$0 < x_j < \infty$, $j=1,\ldots,N$, while for the chiral   
classes, the integration extends over the entire real axis.  
  
For all ten symmetry classes, a change of variables   
$x_j \to \lambda_j$ can be made after which $Z(a)$ can be   
written in the form   
\be  
\label{Z}  
Z\propto \int_{\cal L}d\lambda_1\dots d\lambda_N\,  
\p_{j=1}^Nw(\lambda_j)\,  
\p_{k<j}|\lambda_j-\lambda_k|^{m_o}.  
\e  
The relation between the $\lambda_j$ and the  
$x_j$, the weight function $w(\lambda)$, and  
the integration interval ${\cal L}$ depend on the  
symmetry class and will be specified below.  
  
A general method for calculation of integrals of the   
type (\ref{Z}) is described in chapters 5 and 6 of Mehta's book  
on random  
matrix theory.\cite{Mehta} Here we mention the results  
from Ref.\ \onlinecite{Mehta}, up to a proportionality  
constant that does not depend on $a$.  
For $m_o=2$ the result of integration   
in Eq.~(\ref{Z}) is proportional to the determinant    
\beml  
\label{Z2}  
\beq  
Z&\propto& \det\l[f_{mn}\r]_{m,n=1}^{N},\\  
f_{mn}&=&\int_{\cal L}d\lambda\,w(\lambda)\,p_m(\lambda)p_{n}(\lambda),   
\eq  
\eml  
where the polynomials $\{p_m(\lambda)\}$ form a linear   
independent set of polynomials up to degree $N-1$.  
For $m_o=1$ and even $N$ we have\cite{Mehta}  
\beq\label{Z1even}  
Z&\propto& \sqrt{\det\l[f_{mn}\r]_{m,n=1}^{N}},\\  
f_{mn}&=&\int_{\cal L}d\lambda\,d\lambda'\,   
\epsilon(\lambda-\lambda') w(\lambda)w(\lambda')p_m(\lambda)p_{n}(\lambda'),  
\n   
\eq  
where the sign function $\epsilon(\lambda)$ equals $1$  
for positive $\lambda$ and $-1$ for negative $\lambda$.  
For odd $N$,  
the matrix $f$ must be supplemented   
by an additional row and column  
\beml  
\label{Z1odd}  
\beq  
&&Z\propto \sqrt{\det\l[f_{mn}\r]_{m,n=1}^{N+1}},\\  
&&f_{n,N+1}=-f_{N+1,n}=\int_{\cal L}d\lambda\, w(\lambda)p_n(\lambda).   
\eq  
\eml  
For $m_o=4$, the result is  
\beml  
\label{Z4}  
\beq  
Z&\propto& \sqrt{\det\l[f_{mn}\r]_{m,n=1}^{2N}},\\  
f_{mn}&=&\int_{\cal L}d\lambda\,   
w(\lambda)  
\bigl[p^{\ }_m(\lambda)p'_{n}(\lambda)-p^{\ }_{n}(\lambda)p'_{m}(\lambda)  
\bigr],  
\eq  
\eml  
where the prime stands for the derivative with respect to $\lambda$  
and $\{p_n(\lambda)\}$ form a linear independent set of polynomials   
up to degree $2N-1$.

\subsection{Standard classes}  
  
For the standard universality classes, the partition function  
$Z(a)$ takes the form of the normalization integral of the  
Laguerre ensemble of random matrix theory\cite{Bee96} if we  
set $\lambda=\sinh^2{x}$. In this case, the integration range  
${\cal L} = (0,\infty)$ and the weight function $w(\lambda)$  
is given by $\exp{(-2 a \lambda -a)}$. Integrating   
over the $\lambda_j$, one finds  
that the partition function $Z(a)$  
has a particularly simple dependence on $a$,  
\be  
Z(a)\propto e^{-aN}a^{-\gamma N /2}.  
  \label{eq:Zstandard}  
\e  
Using Eq.\ (\ref{main}), one quickly verifies that this   
corresponds to a constant DOS,  
\be  
\label{constant}  
\nu(\varepsilon) =\nu_{0} \equiv \frac{Nd}{\pi v_F}.  
\e  
(The DOS can acquire a non-universal energy dependence    
on the scales $\varepsilon$ much larger then the inverse  
scattering time $v_F/\ell$,   
where the applicability of the scaling equation breaks down.)  
  
As a matter of fact, for large $a$, one obtains the  
result (\ref{eq:Zstandard}) for all ten symmetry classes,  
as one verifies by substitution $x_j \to a^{-1/2} y_j$ in  
Eq.\ (\ref{eq:Zx}) and subsequent expansion for large $a$.  
This confirms that the DOS $\nu(\varepsilon)$  
approaches the standard value $\nu_0$ for energies far away  
from zero for all symmetry classes.

\subsection{Chiral classes}  
  
For the chiral classes, we arrive at the standard form  
(\ref{Z}) taking   
$\lambda=\exp{(2x)}$, ${\cal L}=(0,\infty)$,  
and  
\be  
w(\lambda)=\lambda^{-\gamma/2}   
\exp{\l[-\case{1}{2}a(\lambda+\lambda^{-1})\r]}.   
\e

When TR is broken, say by a magnetic field ($m_o = 2$;  
class AIII),  
we choose in Eq.~(\ref{Z2}) the polynomials  
$p_n(\lambda)=\lambda^{n-1}$ so that  
\be  
\label{ZAIII}  
Z(a)\propto \det{\l[K_{|m-n|}(a)\r]}_{m,n=1}^{N},  
\e  
where $K_j(a)$ is the modified Bessel function of integer order $j$.  
For small values of the dimensionless imaginary energy  
$a$ one can substitute $K_0(a)\to -\ln{a}$  
and $K_m(a)\to 2^{m-1} a^{-m} (m-1)!$, $m=1,2,\ldots,$ to obtain the  
leading behavior of $Z$ (to logarithmic accuracy).  
For odd $N$ the determinant in Eq.~(\ref{ZAIII})   
is expanded according to  
\be  
\label{Zlog}  
Z(a)\propto a^{-(N^2-1)/2}  
\l[\ln{a}+{\cal O}(1)\r].  
\e  
This gives the asymptotic DOS  
\begin{mathletters}  
\be  
\label{DOSlog}  
\nu(\varepsilon)= \frac{\pi \nu_{0}}  
{|\varepsilon \tau \ln^3{(\varepsilon \tau)}|},  
\qquad 0<\varepsilon\tau\ll 1,  
\e  
where we introduced the time scale   
\begin{equation}  
\tau=N \gamma \ell /v_F.   
\end{equation}  
\end{mathletters}  
It is the time needed to diffuse  
through a wire of length $\sim N\ell$, or,  
equivalently, the mean DOS in a segment of  
the wire of length $N\ell$. (The length scale $N \ell$   
characterizes the crossover between the regimes of  
diffusive dynamics and of localized or critical dynamics.)  
For even $N$ the determinant in Eq.~(\ref{ZAIII})   
is expanded according to  
\be  
Z(a)\propto   
a^{-N^2/2}  
\l[  
1-\case{1}{4}N^2  
a^2\ln^2{a}+{\cal O}(a^2\ln{a})  
\r].  
\e  
This gives rise to a different asymptotic DOS,  
\be  
\label{nuAIIIeven}  
\nu(\varepsilon)=  
\pi \nu_{0}|\varepsilon \tau \ln{(\varepsilon \tau)}|,  
\qquad 0<\varepsilon\tau \ll 1.  
\e  
Observe that the asymptotic results (\ref{DOSlog}) and (\ref{nuAIIIeven})  
do not depend explicitly on $N$   
when the energy is measured in units of $1/\tau$.   
The leading dependence on energy in  
Eqs.~(\ref{DOSlog}) and (\ref{nuAIIIeven})  
has been derived previously in Ref.\ \onlinecite{BMF}  
in an approximation  
that leaves the prefactor $\pi\nu_{0}$ unspecified.

When TR symmetry is present but SR symmetry is broken  
by spin-orbit  
coupling ($m_o=4$, class CII), we choose   
in Eq.~(\ref{Z4})   
the polynomials $p_{2n-1}(\lambda)=\lambda^{2N-n}+\lambda^{n-1}$ and   
$p_{2n}(\lambda)=\lambda^{2N-n}-\lambda^{n-1}$.  
With such a choice the square root of the determinant of  
a $2N\times 2N$ matrix in Eq.~(\ref{Z4})  
reduces to the determinant of a $N\times N$ matrix,   
from which we obtain  
\begin{eqnarray}  
\label{ZCII}  
  Z(a) &\propto& \det{\l[ f_{nm}\r]}_{n,m=1}^{N},  
\\ \n  
  f_{nm} &=& (2N-n-m+1)K_{|n-m|}(a)  
  \nonumber \\ && \mbox{} + (n-m)  
K_{2N-n-m+1}(a).  
\end{eqnarray}  
For odd $N$, the leading term in the asymptotic expansion of   
$Z(a)$ for small $a$ is of the form (\ref{Zlog}),   
hence the small-$\varepsilon$ asymptote of the DOS is   
given by Eq.~(\ref{DOSlog}), just like in the case of  
broken time-reversal symmetry.  
For even $N$ the determinant in Eq.~(\ref{ZCII}) is expanded as  
\be  
Z(a)\propto a^{-N^2}  
\l[1+c_1 a^2+c_2 a^4 \ln^2{a}+{\cal O}(a^4 \ln{a})\r],  
\e  
where  $c_2=\case{1}{48}N^2(N+1)^2$  
and $c_1$ is another $N$-dependent  
coefficient, which does not affect  
the asymptote of the DOS.  
We thus obtain  
\be  
\label{nuCIIeven}  
\nu(\varepsilon)= {\pi \over 3}  
\nu_{0}\l( 1+\case{1}{N}\r)^2  
|(\varepsilon \tau)^3 \ln{(\varepsilon \tau)}|,  
\qquad 0<\varepsilon\tau\ll 1.  
\e  
  
In the presence of both TR and SR symmetry ($m_o=1$, class BDI)  
and  
with the choice of the polynomials $p_n(\lambda)=\lambda^{n-1}$,  
$n=1,\ldots,N$, we obtain for even $N$  
\beq  
\label{ZBDIeven}  
Z(a)&\propto& \sqrt{\det\l[f_{mn}\r]_{m,n=1}^{N}},\\  
f_{mn}&=&\i_0^\infty dx\i_x^\infty dy\,  
(xy)^{-\case{1}{2}(N+1)}e^{-\case{1}{2}a(x+y+x^{-1}+y^{-1})}\n  
\\  
&& \times \l(x^{n-1}y^{m-1}-x^{m-1}y^{n-1}\r).   
\label{ZBDIeven2}  
\eq  
For odd $N$, the antisymmetric matrix $f$ has to be supplemented  
by an additional row and column with  
\begin{eqnarray}  
\label{ZBDIodd}  
f_{n,N+1} &=& \i_0^\infty dx  
x^{-\case{1}{2}(N+1)}e^{-\case{1}{2}a(x+x^{-1})}x^{n-1}  
  \nonumber \\ &=&  
  2 K_{|(N+1)/2-n|}(a).  
\end{eqnarray}  
The analytical continuation $a \to -i \gamma \ell \varepsilon /v_F =  
-i \tau \epsilon/N$ can be done directly in the integrals (\ref{ZBDIeven2}),   
(\ref{ZBDIodd}) if accompanied by an appropriate shift of the   
integration contour of the variables $x$ and $y$ in the complex  
plane. After that, numerical evaluation of   
the partition function (\ref{ZBDIeven})   
for complex $a$ and its derivatives, and hence of the DOS   
$\nu(\epsilon)$, is straightforward.\cite{footchGOE}  
The small-$a$ asymptote of the partition function $Z(a)$ is given  
by Eq.\ (\ref{Zlog}) for odd $N$ and by  
\begin{equation}  
  Z(a) = a^{-(N/2)^2} \left[1 + c a \ln a + {\cal O}(a)\right]  
  \label{eq:ZBDIasymptotEven}  
\end{equation}   
for even $N$,   
where $c = \pi[(N/2-1)!!/(N/2-2)!!]^2$ if $N=4n$  
and $c = 4\pi^{-1} [(N/2-1)!!/(N/2-2)!!]^2$ if $N = 4n-2$,  
$n=1,2,\ldots$.  
As a result, for small energies, the DOS has the form  
(\ref{DOSlog}) for odd $N$, while  
\begin{equation}  
  \nu(\varepsilon) = {\nu_0 c \over N} |\ln (\varepsilon \tau)|,\ \  
  0<\varepsilon \tau \ll 1,  
\end{equation}  
for even $N$, where the coefficient $c$ is given below  
Eq.\ (\ref{eq:ZBDIasymptotEven}). In the limit $N \gg 1$,   
this asymptote simplifies to $\nu(\varepsilon) =  
\nu_0 |\ln(\varepsilon\tau)|$, $0<\varepsilon \tau \ll 1$.

\begin{figure}  
\epsfig{file=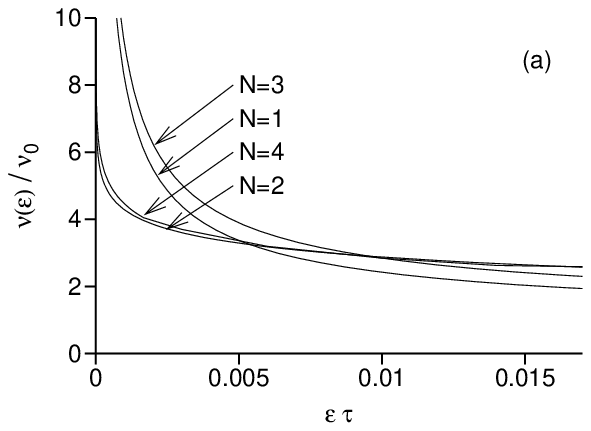,width=8cm}  
\vspace*{0.01cm}  
\epsfig{file=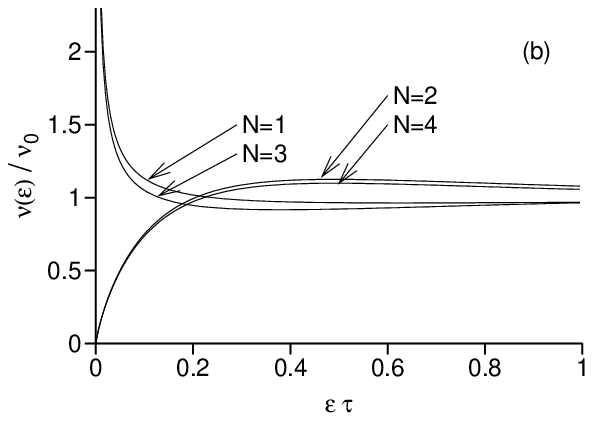,width=8cm}  
\vspace*{0.01cm}  
\epsfig{file=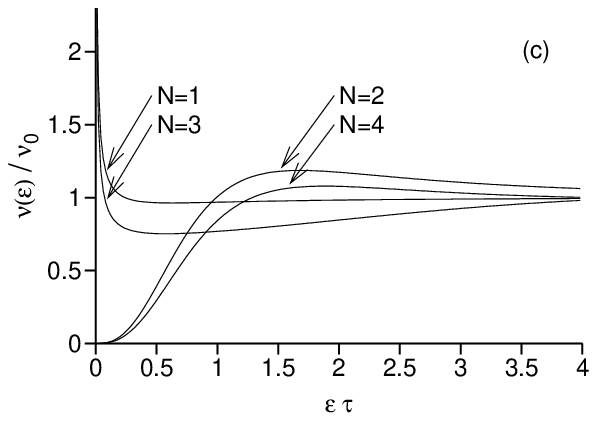,width=8cm}  
\vspace*{0.2cm}  
\caption{%
Density of states for the chiral classes. The energy is  
measured in units of $1/\tau = v_F/(N\gamma\ell)$, the DOS  
is measured in units of $\nu_0 = N d/\pi v_F$.  
Figure\ \protect\ref{fig:1}a is for the case of preserved  
time-reversal symmetry and spin-rotation invariance (Hamiltonian   
of class BDI in Cartan's classification),  
Fig.\ \protect\ref{fig:1}b is for the case of broken time-reversal symmetry  
(class AIII), and Fig.\ \protect\ref{fig:1}c is for the case of preserved  
time-reversal symmetry and broken spin-rotation invariance  
(class CII). The density of states is shown for $N=1$, $2$,  
$3$, and $4$; the values of $N$ are indicated in the figure.  
}  
\label{fig:1}  
\end{figure}

The DOS is plotted for the three chiral classes   
and for various values of $N$   
in Fig.~\ref{fig:1}. The parity effect and the dependence   
on symmetry class is clearly seen.  
    
\subsection{BdG universality classes}  
  
We now apply Eqs.~(\ref{Z2})--(\ref{Z4}) to the BdG   
universality classes. We choose $\lambda=\cosh{2x}$,  
${\cal L}=(1,\infty)$, and  
\be  
w(\lambda)=\exp{(-a \lambda)} (\lambda^2-1)^{(m_l-1)/2},  
\e  
to arrive at the standard form (\ref{Z}). In all   
calculations below we choose the polynomials   
$p_n(\lambda)=\lambda^{n-1}$ to compute $Z(a)$.

When both TR and SR symmetries are present (class CI),   
say as is the case with a dirty $d$-wave superconductor,  
we use Eq.~(\ref{Z2}) and find  
\be  
\label{ZCI}  
Z(a)\propto \det{\l[  
\l(-{d\over da} \r)^{m+n-2}a^{-1}K_1(a)  
\r]}_{m,n=1}^{N}.  
\e   
With the help of the asymptotic limit $a\to0$ of the modified  
Bessel function $K_1$, we obtain  
\be  
Z(a)\propto a^{-N(N+1)}  
\l[  
1+ \case{1}{4}N(N+1) a^2 \ln{a}+{\cal O}(a^2)  
\r].  
\e  
Substituting $a\to -i \tau \varepsilon/N $   
and using Eq.~(\ref{main}) one verifies that the DOS   
vanishes linearly with energy as $\varepsilon\to 0$  
\be  
\nu(\varepsilon)= {\pi\over 2} \nu_{0}  
\l(1+\case{1}{N}\r)|\varepsilon\tau|,  
\qquad 0<\varepsilon\tau\ll1.  
\e  
  
When SR symmetry is present but not TR (class C),  
we find  
\beml  
\label{ZC}  
\beq  
Z(a)&\propto& \sqrt{\det\l[f_{mn}\r]_{m,n=1}^{2N}},\\  
f_{mn}&=&(n-m)\l(-{d \over da} \r)^{n+m-3}{1+a \over a^3}e^{-a} .  
\eq  
\eml  
In the limit $a\to 0$, we obtain  
\be  
\label{ZCexp}  
Z(a)\propto   
a^{-N(2 N+1)}  
\l[  
1+c_1 a^2+ c_2 a^3 +{\cal O}(a^4)  
\r],  
\e  
where $c_2=\case{1}{18}N(N+1)(2N+1)$ and  
$c_1$ is a numerical coefficient which is  
not important for the small energy   
asymptote of the DOS. From  
Eq.\ (\ref{ZCexp}), one verifies that  
the DOS vanishes quadratically with energy as $\varepsilon\to 0$,  
\be  
\nu(\varepsilon)= \nu_{0}\l(1+\case{1}{N}\r)  
\l(1+\case{1}{2N}\r)  
(\varepsilon \tau)^2,  
\qquad 0<\varepsilon\tau\ll 1.  
\e  
  
The linear and quadratic energy dependencies of the DOS  
in the thermodynamic limit and  
for small energies in the classes CI and C, respectively,  
have been found previously by Senthil {\em et al.}\cite{Senthil98,Senthil99}   
The same suppression of the DOS appears in the random  
matrix theory for finite-size systems of classes CI and  
C,\cite{AZ} provided the size of the system is set equal to  
a localization volume (corresponding to a segment of  
length $N \ell$ of the quantum wire).  
  
When TR symmetry is present but not SR (class DIII),  
we find from Eq.~(\ref{Z2})   
\be  
\label{ZDIII}  
Z(a)\propto \det{\l[   
\l(-{d\over da} \r)^{m+n-2}K_0(a)  
\r]}_{m,n=1}^{N}.  
\e   
Note that $N$ is even for the class DIII.  
In the limit $a\to 0$, using the asymptotic  
expansion of the modified Bessel function $K_0$,   
we obtain  
$Z(a)\propto a^{-N(N-1)}[\ln{a}+{\cal O}(1)]$,  
i.e., corresponding to a  
DOS of the form (\ref{DOSlog}).   
For class DIII, the case $N=2$ is special: 
The Lie algebra $so(4,{\Bbb C})$ 
generated by the transfer matrices  
of a quantum wire in class DIII with $N=2$ (see Ref.~\onlinecite{BFGM})  
is not irreducible,  
$so(4,{\Bbb C})\approx sl(2,{\Bbb C})\times sl(2,{\Bbb C})$  
(see Ref.~\onlinecite{Helgason}). 
As a result, in the case DIII with $N=2$  
the scaling of the parameters $x_j$ and of the eigenphases $\phi_j$  
is described by two mean free paths that need not be equal; 
one for each copy of the non-compact Lie algebra $sl(2,{\Bbb R})$. 
The scaling equations (\ref{eq:PL}), (\ref{FP}),  
and (\ref{eq:FPphiall}), in which only one mean free path appears,  
correspond to the special case where these two mean free paths  
are equal, which   
requires fine tuning of the disorder. Only in that case the DOS  
singularity (\ref{DOSlog}) is found. In Refs.\    
\onlinecite{Motrunich00} and \onlinecite{GruzbergPriv}  
it is shown that the general case of different mean free paths has a  
power law dependence of the DOS for $\varepsilon \to  
0$ with a non-universal exponent.

\begin{figure}  
\epsfig{file=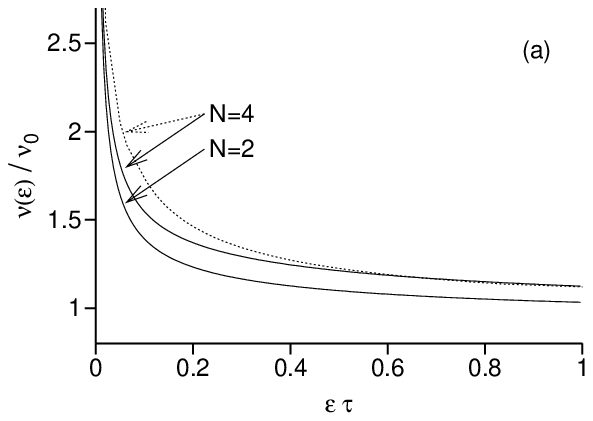,width=8cm}  
\vspace*{0.01cm}  
\epsfig{file=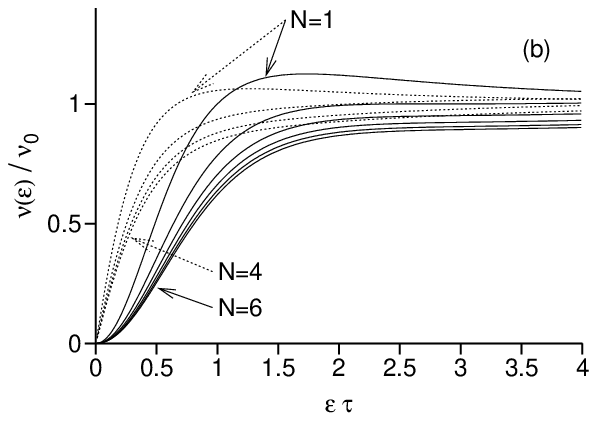,width=8cm}  
\vspace*{0.2cm}  
\caption{%
Density of states for the BdG universality classes.  
Fig.\ \protect\ref{fig:2}a contains results for  
classes DIII (solid curve) and D (dotted curve),  
for which spin-rotation invariance is broken;  
Fig.\ \protect\ref{fig:2}b is for classes C (solid)  
and CI (dotted), where spin-rotation  
invariance is preserved.   
The channel number $N$ is indicated in the figure.  
}  
\label{fig:2}  
\end{figure}  
  
Finally,  
when neither SR nor TR symmetries are present (class D),  
it is enough to use only Eq.~(\ref{Z1even}) because   
the number of eigenvalues $N$ is a multiple of four   
in this case.   
We then obtain  
\beq  
\label{ZD}  
Z(a)&\propto& \sqrt{\det\l[ f_{mn} \r]_{m,n=1}^{N}},\\  
f_{mn}&=&\i_1^\infty dx\i_x^\infty dy\,  
\frac{x^{n-1}y^{m-1}-x^{m-1}y^{n-1}  
}{\sqrt{(x^2-1)(y^2-1)}}\, e^{-a(x+y)}. \n  
\eq  
With the help of the asymptotic limit $a\to 0$   
we find Eq.\ (\ref{DOSlog})   
for the asymptotic DOS. Numerical evaluation  
of the DOS is possible using (\ref{ZD}).   
The substitution   
$a\to-i\tau\varepsilon/N$ can be done  
before the integration and the   
convergency of the integrals has to be  
provided by the appropriate shift of the   
integration contour of $x$ and $y$ in the  
complex plane.   
A divergence of the DOS was also predicted  
by Senthil and Fisher for two-dimensional systems of class   
D/DIII.\cite{Senthil00}  
  
The case $N=2$, which corresponds to a ``spinless'' class D and was  
studied in Refs.\ \onlinecite{Motrunich00} and \onlinecite{GruzbergPriv}, 
is again very special 
[see the discussion of class DIII with $N=2$ below Eq.~(\ref{ZDIII})]:   
The Lie algebra $so(2,2)$ generated by the transfer matrices  
of a quantum wire in class D with $N=2$ (see Ref.~\onlinecite{BFGM})  
is not irreducible, 
$so(2,2)\approx sl(2,{\Bbb R})\times sl(2,{\Bbb R})$ 
(see Ref.\ \onlinecite{Helgason}). 
Therefore, as is explained in Refs.\ \onlinecite{Motrunich00} and  
\onlinecite{GruzbergPriv},  
instead of the Dyson singularity (\ref{DOSlog}) an algebraic  
$\varepsilon$-dependence with non-universal exponent is  
obtained for the DOS singularity at $\varepsilon=0$ 
whenever the two independent mean free paths associated to the two copies 
of the non-compact Lie Algebra $sl(2,{\Bbb R})$ are unequal. 
  
The DOS is shown for the four BdG classes   
and for various $N$ in Fig.~\ref{fig:2}.   
Note that there is no parity effect for the   
BdG classes and the DOS becomes almost independent of $N$  
beyond $N\approx 4$, provided energy is measured in  
units of $1/\tau$.  
  
\section{Conclusion}

In this paper, we have presented a unified picture for both   
transport and the density of states (DOS) in a long  
quantum wire, for all ten (pure) Cartan symmetry classes.\cite{foot2}  
While it was well known that the dependence of the conductance on  
the length $L$ of the quantum wire could be inferred from the  
theory of diffusion on symmetric spaces through the   
Dorokhov-Mello-Pereyra-Kumar (DMPK) equation and its   
generalizations,\cite{Caselle,Hueffmann,Beenakker}   
we have shown that the same geometrical framework underlies the   
dependence of the DOS on the energy $\varepsilon$ (measured  
with respect to the band center or Fermi energy $\varepsilon=0$).  
Our unified picture consists of a general scheme that permits   
the construction of  
scaling equations for the eigenvalues $e^{2 i \phi_j}$ of the  
matrix product $r^{\dagger}(-\varepsilon) r(\varepsilon)$ of  
reflection matrices $r$ at different energies for a  
disordered quantum wire and for all  
ten pure Cartan symmetry classes. [Here $j=1,\ldots,N$, where $N$   
is the number of independent  
eigenvalues of $r^{\dagger}(-\varepsilon) r(\varepsilon)$.]  
Using the analytical   
continuation $\varepsilon \to i \omega$, this scaling equation  
could be obtained from that for the reflection  
eigenvalues $R_j$ of the quantum wire in the presence of absorption,  
which, in turn, can be obtained from the DMPK equation.   
Using the ``node-counting theorem'' that relates the $L$-dependence   
of the eigenphases $\phi_j$ and the DOS,\cite{Schmidt,But93}   
we were thus able to find the DOS in the  
thermodynamic limit $L \to \infty$ from the stationary solution  
of the scaling equation.  
  
Relatively little attention has been paid to the behavior of the   
density of states (DOS) in the standard problem of Anderson   
localization for a system of infinite size, since this   
DOS,  
with the exception of its tails, is rather insensitive to disorder.   
This is not true in the presence of symmetries for which   
one energy in the spectrum plays a special role, as is the case for   
the chiral and Bogoliubov-de-Gennes (BdG) universality   
classes, where a sublattice symmetry or the presence of a   
superconducting order parameter singles out the band center or   
the Fermi energy $\varepsilon=0$, respectively.   
In those cases, the DOS is singular around zero energy,   
the energy scale for the singularity being $\hbar/\tau$, where,  
for the case of a long quantum wire we consider in this paper,   
$\tau \sim N^2 \ell/v_F$ is the time for diffusion through a   
segment of the wire of length $N \ell$. The precise form of the   
singularity  
depends on the symmetry class and, for the chiral classes, on the  
parity of $N$. The results are summarized in Table \ref{tab:2} for  
the limit of a multichannel quantum wire, $N \gg 1$. It is in this  
limit, that a universal dependence of the DOS on the symmetry class  
is expected.   
  
One might be tempted to argue that, for those classes where the  
quasiparticle states are localized at and near zero energy,  
the singular behavior of the DOS in the thermodynamic limit  
is controlled by the (zero-dimensional) random matrix theory   
for an effective finite-size system with the linear size of the   
localization length $\xi = N \ell$.  
Whereas this argument predicts the correct energy scale   
$\hbar/\tau$ for the  
singularity, it fails  
to reproduce the extra logarithm in the  
functional form of the singularity   
for the chiral classes, see Table \ref{tab:2}.  
The origin of the logarithm in the DOS can be found in the  
presence of two competing localized modes in that case,   
that each have the  
same localization length, see Refs.\ \onlinecite{BMSA,BMF}.  
As was shown in Ref.\ \onlinecite{BMF}, the presence of a small  
amount of dimerization of the hopping amplitudes lifts the  
degeneracy of the localization lengths and removes the logarithm  
from the functional form of the DOS singularity. On the other  
hand, for the  
BdG classes C and CI, there is only one localized mode,\cite{BFGM}  
which explains the absence of the extra logarithm   
in the functional form of the DOS in those classes.  
  
\begin{table}  
\begin{tabular}{l|l|l|l}  
class & TR & SR & $\nu(\varepsilon)$ for $0<\varepsilon \tau \ll 1$ \\  \hline   
& Yes & Yes & $\nu_0 |\ln(\varepsilon \tau)|$ \\ chiral,  
$N$ even & No & & $\pi \nu_0 |\varepsilon \tau \ln(\varepsilon \tau)|$ \\  
 & Yes & No & $(\pi \nu_0/3) |(\varepsilon \tau)^3  
   \ln(\varepsilon \tau)|$ \\ \hline   
chiral, $N$ odd  
& & & $\pi \nu_0/|\varepsilon \tau \ln^3 (\varepsilon \tau)|$  
\\ \hline  
& Yes & Yes & $(\pi \nu_0/2) |\varepsilon \tau|$ \\  
BdG & No & Yes & $\nu_0 |\varepsilon \tau|^2$ \\  
& Yes & No & $\pi \nu_0/|\varepsilon \tau \ln^3 (\varepsilon \tau)|$ \\  
& No  & No & $\pi \nu_0/|\varepsilon \tau \ln^3 (\varepsilon \tau)|$   
\end{tabular}\medskip  
  
\caption{\label{tab:2} Asymptotes of the DOS   
$\nu(\varepsilon)$ for small energies $0<\varepsilon \tau \ll 1$  
and for $N \gg 1$,  
where $\tau = N \gamma \ell/v_F$, $\gamma$ is given in Eqs.\  
(\protect\ref{eq:gammaBdG}) and (\protect\ref{eq:gammachiral}),   
$\ell$ is the mean free path, and  
$v_F$ the Fermi velocity. (Expressions for the DOS asymptotes for  
finite  
$N$ can be found in the text.)}  
\end{table}  
  
\acknowledgements  
We thank C. W. J. Beenakker, K. Damle, and I. A. Gruzberg for discussions.   
This work was supported by the  
Dutch Science Foundation NWO/FOM   
and by INTAS Grant No.\ 97-1342 (MT), the Sloan foundation (PWB),  
and by the Grant-in-Aid for Scientific Research on Priority Areas (A)  
from the Ministry of Education, Science, Sports and Culture of  
Japan No.\ 12046238 (AF).  
PWB gratefully acknowledges the  
hospitality of the Instituut Lorentz of Leiden University where  
part of this work was done. Upon completion of this manuscript, we  
learned of a related paper by Motrunich {\em et al.} (Ref.\  
\onlinecite{Motrunich00}) on  
single-channel quantum wires with BdG symmetry and broken  
spin-rotation invariance, where the same conclusion was reached  
with respect as to the DOS singularity for the pure Cartan  
symmetry classes D and DIII.   
  
  


\begin{references}
\bibitem{foot}  
        In the standard and chiral classes, the symmetry  
        classes with and without spin-rotation invariance coincide  
        once time-reversal symmetry is broken.
\bibitem{exception}  
        The scaling equation is not generic for Cartan's class  
        DIII if $N=2$, see the discussion following Eq.~(\ref{ZDIII}).
\bibitem{footmo}  
The multiplicity $m_o$ is known as the Dyson symmetry   
parameter $\beta$  
for the standard ensembles from random matrix theory,  
see Ref.~\onlinecite{Mehta}.
\bibitem{foot2}  
The issue whether a generic disordered electron system belongs to  
one of Cartan's classes, or whether it requires fine tuning of  
the disorder distribution for a system to fall into one of the  
ten symmetry classes has been the subject of recent controversy  
in the literature (see Refs.~\onlinecite{Motrunich00}, \onlinecite{rbim},   
and references therein). We do not want to add to this debate here.  
Instead, in this paper, our starting point is the assumption that  
the physical system at hand has a Hamiltonian belonging to one of  
the pure Cartan symmetry classes from Table \ref{tab:1}, for which  
the scaling equation for the transmission eigenvalues is given by  
Eq.~(\ref{eq:PL}), and our  
goal is to study the DOS singularity near zero energy  
under that assumption. With ``pure'' Cartan classes, we refer to  
those quantum wires whose Hamiltonian is taken from an 
ensemble of which the distribution is as random as 
possible within the bounds set by symmetry, or, 
alternatively, for which the motion of the transfer matrix  
in the appropriate symmetric space is isotropic.\cite{Caselle,Hueffmann}   
Examples of Hamiltonians for the pure Cartan classes can be  
found in Refs.~\onlinecite{BMSA} and \onlinecite{BFGM}.
\bibitem{footst} One reason why the stationary solution  
$P_{\varepsilon;{\rm st}}$ is not of the simple form (\ref{solution})  
is that the phases $\phi_j$ take values on the unit circle, which is  
not simply connected.
\bibitem{footchGOE}  
With the choice of the polynomials   
$p_{2n-1}(\lambda)=\lambda^{2N-n}+\lambda^{n-1}$ and   
$p_{2n}(\lambda)=\lambda^{2N-n}-\lambda^{n-1}$  
the partition function can be written as a determinant of  
a matrix of size $N/2$ [$(N+1)/2$ for odd $N$].   
Then, the integration can be done analytically for $N=2$ and   
$N=3$   
\begin{eqnarray*}  
\l.Z(a)\r|_{N=2}&\propto& \frac{1}{a}\int_{2a}^\infty  
K_0(t)\, dt, \\  
\l.Z(a)\r|_{N=3}&\propto&   
\frac{1}{a}  
\l[  
K_0(a)K_1(2a)-K_1(a)K_0(2a)  
\r].  
\end{eqnarray*}
\end{references}
\end{document}